\documentclass[a4paper,10pt,twoside,preprint,aps ,nofootinbib]{cpc-hepnp}

\usepackage{graphicx}
\pdfoutput=1
\usepackage{epsfig}
\usepackage{amsmath}
\usepackage{amsfonts}
\usepackage{amssymb}
\usepackage{color}
\usepackage{dcolumn}
\usepackage{hyperref}
\usepackage{multirow}
\usepackage{CJKutf8}
\usepackage{array}
\usepackage{siunitx}
\usepackage{booktabs}

\providecommand{\U}[1]{\protect\rule{.1in}{.1in}}

\newcommand{\f}{\begin{equation}}
\newcommand{\ff}{\end{equation}}
\newcommand{\fa}{\begin{eqnarray}}
\newcommand{\ffa}{\end{eqnarray}}

\begin{document}

\title{\boldmath The implications of gamma-ray photons from LHAASO on Lorentz symmetry
\thanks{We are very grateful to Zhe Chang, Yu-Chen Ding, Chao-guang Huang  and Qinghua Zhu for helpful discussions on LHAASO project and relevant topics. We would also like to thank the anonymous referees for helpful comments and suggestions. This work is supported in part by the Natural Science Foundation of China under
Grant No.~11875053 and 12035016. Liu Yuxuan acknowledges the
support from the National Postdoctoral Program for Innovative
Talents BX2021303, funded by China Postdoctoral Science
Foundation.}}

\author{Yi Ling  $^{1,2}$ \email{lingy@ihep.ac.cn}
\quad Yuxuan Liu $^{1,2,3}$ \email{liuyuxuan@ucas.ac.cn}
\quad Sai Wang $^{1,2}$ \email{wangsai@ihep.ac.cn}
\quad Meng-He Wu  $^{1,2}$ \email{mhwu@ihep.ac.cn}}

\maketitle

\address{
$^1$ Institute of High Energy Physics, Chinese Academy of Sciences, Beijing, 100049, China \\
$^2$ School of Physics, University of Chinese Academy of Sciences, Beijing, 100049, China \\
$^3$ Kavli Institute for Theoretical Sciences (KITS), University of Chinese Academy of Sciences,
Beijing 100190, China }
\maketitle

\begin{abstract}
The Large High Altitude Air Shower Observatory (LHAASO) has
reported the measurement of photons with high energy up to 1.42 PeV
from twelve gamma-ray sources. We are concerned with the implications
of LHAASO data on the fate of Lorenz symmetry at such high energy
level, thus we consider the interaction of the gamma ray with
those photons in cosmic microwave background (CMB), and compute
the optical depth, the mean free path as well as the survival
probability for photons from all these gamma-ray sources.
Employing the threshold value predicted by the standard special
relativity, it is found that the lowest survival probability for
observed gamma ray photons is about 0.60, which is a fairly high
value and implies that abundant photons with energy above the
threshold value may reach the Earth without Lorentz symmetry
violation. We conclude that it is still far to argue that the
Lorentz symmetry would be violated due to the present observations
from LHAASO.

\end{abstract}

\section{Introduction}

The recent observation by the LHAASO cooperative team has opened a
new window for exploring the ultra-high-energy (UHE) cosmic rays and
challenged our current understanding on the origin of such
UHE gamma radiation as well as the theory of
high energy physics. The LHAASO experiment has detected a large
number of gamma-ray photons with energy exceeding 100 TeV from twelve
gamma-ray sources in our galaxy\cite{Cao}. The highest energy up
to 1.42 PeV announces the coming of the new age of Galactic
PeVatrons in very-high-energy astronomy. Before this the photon
spectra of galactic sources with energy beyond 100 TeV have been
detected by Tibet-AS$\gamma$
\cite{Amenomori:2019rjd,TibetASgamma:2021tpz},
HAWC\cite{HAWC:2019xhp,HAWC:2019tcx} as well as
Carpet-2\cite{Carpet-3:2021omd}.

In theory, the observation of UHE cosmic ray brings two
fundamental problems. One is on the origin of those UHE gamma
radiation\cite{Hillas1984}. It remains mysterious what kind of
galactic sources could provide such extremal physical conditions
to accelerate particles to such high energy
level\cite{Aloisio2018}. The other is on the propagation of UHE
cosmic ray in the universe or galaxy, which involves in  various
interactions between the cosmic ray and the background such as CMB
or interstellar radiation fields (ISRF). One outstanding problem
is on the fate of Lorentz symmetry at such high energy level,
which determines the threshold value of the possible interactions
and leads to different energy cutoffs that we could observe in the
Earth. It has been extensively investigated in literature on the
possibility of treating the UHE cosmic ray as the probe of Lorentz symmetry
violation\cite{Maccione:2010sv,Liberati:2013xla,Vasileiou:2013vra,Mavromatos:2010pk,Shao:2010wk},
which, on the theoretical side, may be treated as the signal of
quantum gravity
effects\cite{Amelino-Camelia:1996bln,Amelino-Camelia:1998bln,Mattingly:2005re,Jacobson:2005bg,Amelino-Camelia:2008aez,Liberati:2009pf}.

The discussion on Lorentz symmetry violation based on LHAASO data
has been presented in
\cite{Chen:2021hen,Li:2021cdz,Li:2021tcw,Satunin:2021vfx,LHAASO:2021opi}. The constraints on
the mass scale of Lorentz symmetry violation has been discussed
based on the LHAASO data. In particular, the Lorentz violation due
to the subluminal scenario seems to be preferred and the
constraints on the mass scale becomes stronger in comparison with
those appeared in literature \cite{Martinez-Huerta:2017ulw,Astapov:2019xmt,Satunin:2019gsl,HAWC:2019gui,Wei:2021ite}. One of the main reasons to consider
the Lorentz symmetry violation comes from the fact that the
observed energy of gamma ray photons is much higher than the
threshold value of pair-production interaction. In the universe,
high-energy cosmic rays interact with the background. One typical
process for gamma rays is the interaction with the CMB photons,
which is $\gamma \gamma \rightarrow e^{+} e^{-}$. From the
threshold theorem in standard special relativity, it is
straightforward to obtain the threshold energy for gamma-ray
photons, which reads as $m^2_e/\epsilon_{b}$, with $\epsilon_{b}$
being the energy of background photons and $m_e$ being the mass of electron. Substituting the energy of
CMB photons into this expression, one finds the threshold energy
for gamma ray photons is about $400$ TeV. Since the energy of some
gamma-ray photons detected by LHAASO is much higher than this
threshold value, apparently it requires people to increase the
theoretical value of threshold energy by modifying the ordinary
dispersion relations in special relativity, thus opens a window
for the possibility of Lorentz symmetry violation, which has been
investigated in Ref.\cite{Li:2021tcw}. On the other hand, inspired by \cite{Li:2021tcw}, we intend to understand the LHAASO data from an alternative point
of view. Although the energy of gamma-ray photons detected by
LHAASO is much higher than the threshold value of the pair
production, which means the gamma-ray photons must interact with
background photons and thus decay during the propagation, the key
point is how many photons would survive during the propagation and
finally reach the Earth, which obviously depends on the number
density of background photons as well as the distance between the
gamma-ray sources and the Earth. This problem is addressed by
considering the transparency of the universe to the gamma
rays\cite{Gould:1967zza,Moskalenko:2005ng,Angelis,GuedesLang:2017sfl}.
The averaged distance that the gamma rays can  propagate through
the background is described by the optical depth $\tau$. Roughly
speaking, if the optical depth $\tau$ is less than one, then the
gamma ray can penetrate the medium between the source and the
Earth.

As a matter of fact, the attenuation of galactic gamma-rays due to
the interaction with photons from CMB and ISRF is briefly
discussed in the original paper by LHAASO team and the opacity for
four gamma sources is presented in its Figure 6\cite{Cao}. In this
note we are very concerned with the fate of Lorentz symmetry, thus
we wonder what is the {\it lowest} value for the survival
probability among the eleven gamma ray sources
\footnote{Since no possible origin was found for the source
``LHAASO J2108+5157'' in \cite{Cao}, we only analyze eleven
of total twelve gamma ray sources.}. Therefore, following the
suggestion from \cite{Li:2021tcw}, we intend to elaborate the
investigation on gamma ray photons with the background based on
LHAASO data. Specifically, we will compute the optical depth, the
mean free path as well as the survival probability for photons
from all gamma-ray sources in LHAASO, and then find the lowest
survival probability for observed gamma ray photons. Our key
result is that among  the eleven gamma ray sources, the optical
depth is always less than one, and the lowest survival probability
is about 0.60, which is a fairly high value and implies that
abundant photons with energy above the threshold value may reach
the Earth without Lorentz symmetry violation. We conclude that it
is still far to argue that the Lorentz symmetry would be violated
due to the present observations from LHAASO.

\section{The optical depth and the survival probability of gamma ray photons }

In this section we just present the main process for computing the
optical depth and the survival probability of gamma ray photons,
and the detailed derivation and discussion can be found in
Ref. \cite{Angelis}. Since the photons detected by the LHAASO are
galactic and in the range of 100 TeV to 10 PeV, we only consider
the interaction process with CMB photons, which is dominant in
comparison with the process with ISRF photons, as shown in Ref.
\cite{Angelis,GuedesLang:2017sfl}.

Usually, during a propagating
process, the survival probability of photons is defined as
\begin{equation}
P_{\gamma \rightarrow \gamma}\left(E_{0},
z_{s}\right)=e^{-\tau_{\gamma}\left(E_{0}, z_{s}\right)},
\end{equation}
where $E_{0}$ is the observed energy and $z_{s}$ is the redshift.
The key quantity $\tau_{\gamma}\left(E_{0}, z_{s}\right)$ is the
optical depth which characterizes the dimming of the source at
$z_{s}$. During the propagation in the universe,
$\tau_{\gamma}\left(E_{0}, z_{s}\right)$ is given
by\cite{Angelis,Gould:1967zza}
\begin{equation}\label{depth}
\begin{aligned}
&\tau_{\gamma}\left(E_{0}, z_{s}\right)=\int_{0}^{z_{s}} \mathrm{~d} z \frac{\mathrm{d} l(z)}{\mathrm{d} z} \int_{-1}^{1} \mathrm{~d}(\cos \varphi) \frac{1-\cos \varphi}{2}  \\
&\times \int_{\epsilon_{\mathrm{thr}}(E(z), \varphi)}^{\infty}
\mathrm{d} \epsilon(z) n_{\gamma}(\epsilon(z), z) \sigma_{\gamma
\gamma}(E(z), \epsilon(z), \varphi),
\end{aligned}
\end{equation}
where $\varphi$ is the scattering angle, and $n_{\gamma}$ is the number density of background photons. $\sigma_{\gamma \gamma}$ is the cross-section of the interaction of pair production and $\epsilon_{\mathrm{thr}}$ is the threshold energy of background photons in the interaction, while $\epsilon(z)$ and $E(z)$ are the energy of background photons and gamma ray photons at a certain redshift $z$, respectively. In standard special relativity with
Lorentz symmetry, it can be derived that
$\epsilon_{\mathrm{thr}}(E, \varphi) = \frac{2 m_{e}^{2}
c^{4}}{E(1-\cos \varphi)}$. In addition, $\mathrm{d} l(z)/\mathrm{d} z$ is the distance travelled by a photon per unit redshift at redshift
  $z$, which within the standard $\Lambda $CDM cosmological model is
  given by
  \begin{equation}
  \frac{\mathrm{d} l(z)}{\mathrm{d} z}=\frac{c}{H_{0}}
  \frac{1}{(1+z)\left[\Omega_{\Lambda}+\Omega_{M}(1+z)^{3}\right]^{1
  / 2}},
  \end{equation}
  where $H_{0}\simeq 7 \times 10^3 \mathrm{cm} \  \mathrm{s^{-1} \
  kpc}^{-1} $ is the Hubble-Lemaitre constant, and
  $\Omega_{\Lambda}\simeq0.7$ is the dark energy density, and
  $\Omega_{M}\simeq 0.3$ is the matter energy density.

As mentioned above, for galactic sources as presented in Ref. \cite{Cao}, the background photons are dominated by CMB photons and the effect of the redshift $z$ on the quantities in (\ref{depth}) is ignored throughout the paper, since the redshift is tiny. In this context, the number density of CMB photons $ n_{\gamma}(\epsilon(z), z)$ can be
approximately written as
\begin{equation}
\begin{aligned}
&n_{\gamma}(\epsilon)=\frac{8 \pi  \epsilon ^2}{c^3 h^3
\left(e^{\frac{\epsilon }{k T}}-1\right)},
\end{aligned}
\end{equation}
where $k$ is Boltzmann constant and $T$ is the temperature of the background. While
the cross-section $\sigma_{\gamma \gamma}(E(z), \epsilon(z), \varphi)$ is given by \cite{Angelis,GuedesLang:2017sfl}
\begin{equation}
\begin{aligned}
\sigma_{\gamma \gamma}(E, \epsilon, \varphi) =\frac{2 \pi
\alpha^{2}}{3 m_{e}^{2}} W(\beta)  \simeq 1.25 \cdot 10^{-25}
W(\beta) \; \mathrm{cm}^{2},
\end{aligned}
\end{equation}
with
\begin{equation}
\begin{aligned}
&W(\beta)=\left(1-\beta^{2}\right)\left[2
\beta\left(\beta^{2}-2\right)+\left(3-\beta^{4}\right) \ln
\left(\frac{1+\beta}{1-\beta}\right)\right],
\end{aligned}
\end{equation}
where $\alpha$ is the fine-structure constant,  and $\beta=(1-\epsilon_{\mathrm{thr}}/\epsilon)^{1/2}$.

Furthermore, the distance $D$ is  more appropriate than the redshift $z_{s}$ for galactic sources and their relationship is given by
\begin{equation}
\begin{aligned}
&D= c z_{s}/H_0.
\end{aligned}
\end{equation}
Thus to the leading order of $D$,  Eq.(\ref{depth}) is replaced by the following expression
\begin{equation}\label{eq_od}
  \tau_{\gamma}\left(E_0, D\right)=D
  \int_{-1}^{1} \mathrm{~d}(\cos \varphi) \frac{1-\cos \varphi}{2} \int_{\epsilon_{\mathrm{thr}}(E, \varphi)}^{\infty}
  \mathrm{d} \epsilon \; n_{\gamma}(\epsilon) \sigma_{\gamma
  \gamma}(E, \epsilon, \varphi),
  \end{equation}
which is the key formula used in this paper. Also, notice that for the source distance corresponding to the tiny redshift, we have $E_0\approx E$.

Once the optical depth is computed, one can obtain the mean
free path of gamma ray photons by the following
relation\cite{Angelis},
\begin{equation}\label{eq_mfp}
\lambda_{\gamma}(E_0,D)=\frac{D}{\tau_{\gamma}(E_0,D)},
\end{equation}
where $\lambda_{\gamma}(E_0,D)$ stands for the mean free path of
photons with energy $E_0$.

\begin{table*}[t]
\newcommand{\tabincell}[2]{\begin{tabular}{@{}#1@{}}#2\end{tabular}}
\caption{\label{Tabel1} The optical depth and the survival probability of gamma ray photons detected by LHAASO. }
\begin{center}
\begin{tabular}{ccccccc}
\toprule
LHAASO Source & \tabincell{c}{ Distance \\ (kpc) }  & \tabincell{c}{Observed energy \\ (PeV)}& Optical depth & \tabincell{c}{ Mean free \\ path (kpc)} & \tabincell{c}{Survival\\ probability } \\ \hline

LHAASO  J2032+4102 & 1.40 $\pm 0.08$ & 1.42 $\pm 0.13$  & $0.18_{-0.02}^{+0.02}$ & $7.68_{-0.19}^{+0.25}$ & $0.83_{-0.01}^{+0.01}$   \\ \hline

LHAASO  J0534+2202 & 2.0 & 0.88 $\pm 0.11$  & $0.20_{-0.02}^{+0.02}$ & $10.00_{-0.87}^{+1.30}$ & $0.82_{-0.02}^{+0.02}$   \\  \hline

LHAASO  J1825-1326 & 3.1$\pm 0.2$ & 0.42 $\pm 0.16$ &  $0.11_{-0.08}^{+0.10}$ &  $29.07_{-13.03}^{+91.85}$ & $0.90_{-0.08}^{+0.08}$   \\
                   & 1.6 & 0.42 $\pm 0.16$ & $0.06_{-0.04}^{+0.04}$ &  $29.07_{-13.03}^{+91.85}$ & $0.95_{-0.04}^{+0.04}$   \\ \hline

LHAASO  J1839-0545 & 4.8 & 0.21 $\pm 0.05$  & $0.02_{-0.01}^{+0.02}$ & $305.60_{-184.67}^{+1116.01}$ & $0.98_{-0.02}^{+0.01}$   \\
                   & 1.3 & 0.21 $\pm 0.05$ &  $0.00_{-0.00}^{+0.01}$ & $305.60_{-184.67}^{+1116.01}$ & $1.00_{-0.01}^{+0.00}$  \\ \hline

LHAASO  J1843-0338 & 9.6$\pm 0.3$ &  $0.26_{-0.10}^{+0.16}$ & $0.08_{-0.07}^{+0.26}$ & $120.92_{-91.85}^{+1300.68}$ &   $0.92_{-0.21}^{+0.07}$   \\ \hline

LHAASO  J1849-0003 & 7 & 0.35 $\pm 0.07$ & $0.15_{-0.08}^{+0.09}$ & $45.71_{-16.64}^{+46.23}$ &  $ 0.86_{-0.07}^{+0.07}$  \\
                   & 5.5 & 0.35 $\pm 0.07$ & $0.12_{-0.06}^{+0.07}$ & $45.71_{-16.64}^{+46.23}$ &   $0.89_{-0.06}^{+0.06}$  \\ \hline

LHAASO  J1908+0621 & 2.4 & 0.44 $\pm 0.05$ & $0.09_{-0.02}^{+0.02}$ & $26.28_{-4.98}^{+8.25}$&   $0.91_{-0.02}^{+0.02}$   \\
                   & 3.4 & 0.44 $\pm 0.05$ &  $0.13_{-0.03}^{+0.03}$ &$26.28_{-4.98}^{+8.25}$ &  $0.88_{-0.03}^{+0.03}$  \\ \hline

LHAASO  J1929+1745 & 4.6 &  $0.71_{-0.07}^{+0.16}$ & $0.37_{-0.04}^{+0.08}$& $12.31_{-2.21}^{+1.67}$&  $0.69_{-0.05}^{+0.03}$ \\
                   & 6.2 & $0.71_{-0.07}^{+0.16}$ & $0.50_{-0.06}^{+0.11}$&$12.31_{-2.21}^{+1.67}$ &   $0.60_{-0.06}^{+0.04}$ \\
                   & $6.3_{-0.7}^{+0.8}$ & $0.71_{-0.07}^{+0.16}$ & $0.51_{-0.11}^{+0.19}$ &$12.31_{-2.21}^{+1.67}$ &    $0.60_{-0.10}^{+0.07}$  \\ \hline

LHAASO  J1956+2845 & 2 & 0.42 $\pm 0.03$ & $0.04_{-0.01}^{+0.01}$ &$29.07_{-3.99}^{+5.45}$&$0.93_{-0.01}^{+0.01}$\\
                   & 2.3$\pm 0.2$ & 0.42 $\pm 0.03$ &  $0.08_{-0.02}^{+0.02}$ & $29.07_{-3.99}^{+5.45}$ &  $0.92_{-0.02}^{+0.02}$  \\ \hline

LHAASO  J2018+3651 & $1.8_{-1.4}^{+1.7}$ & 0.27 $\pm 0.02$ & $0.02_{-0.01}^{+0.03}$ &$104.88_{-23.49}^{+36.22}$ &  $0.98_{-0.03}^{+0.01}$  \\
                   & 3.3$\pm 0.3$ & 0.27 $\pm 0.02$ & $0.03_{-0.01}^{+0.01}$ & $104.88_{-23.49}^{+36.22}$ &    $0.97_{-0.01}^{+0.01}$ \\
                   & 4.0$\pm 0.5$   & 0.27 $\pm 0.02$ & $ 0.04_{-0.01}^{+0.02}$ &$104.88_{-23.49}^{+36.22}$ & $0.96_{-0.02}^{+0.01}$ \\ \hline

LHAASO  J2226+6057 & 0.8 & 0.57 $\pm 0.19$ & $0.05_{-0.03}^{+0.02}$ & $16.48_{-5.03}^{+20.37}$ & $0.95_{-0.02}^{+0.03}$   \\
\bottomrule
\end{tabular}
\end{center}
\end{table*}

\section{Numerical Results}\label{sec3}

\begin{figure} [h]
  \center{
  \includegraphics[scale=0.25]{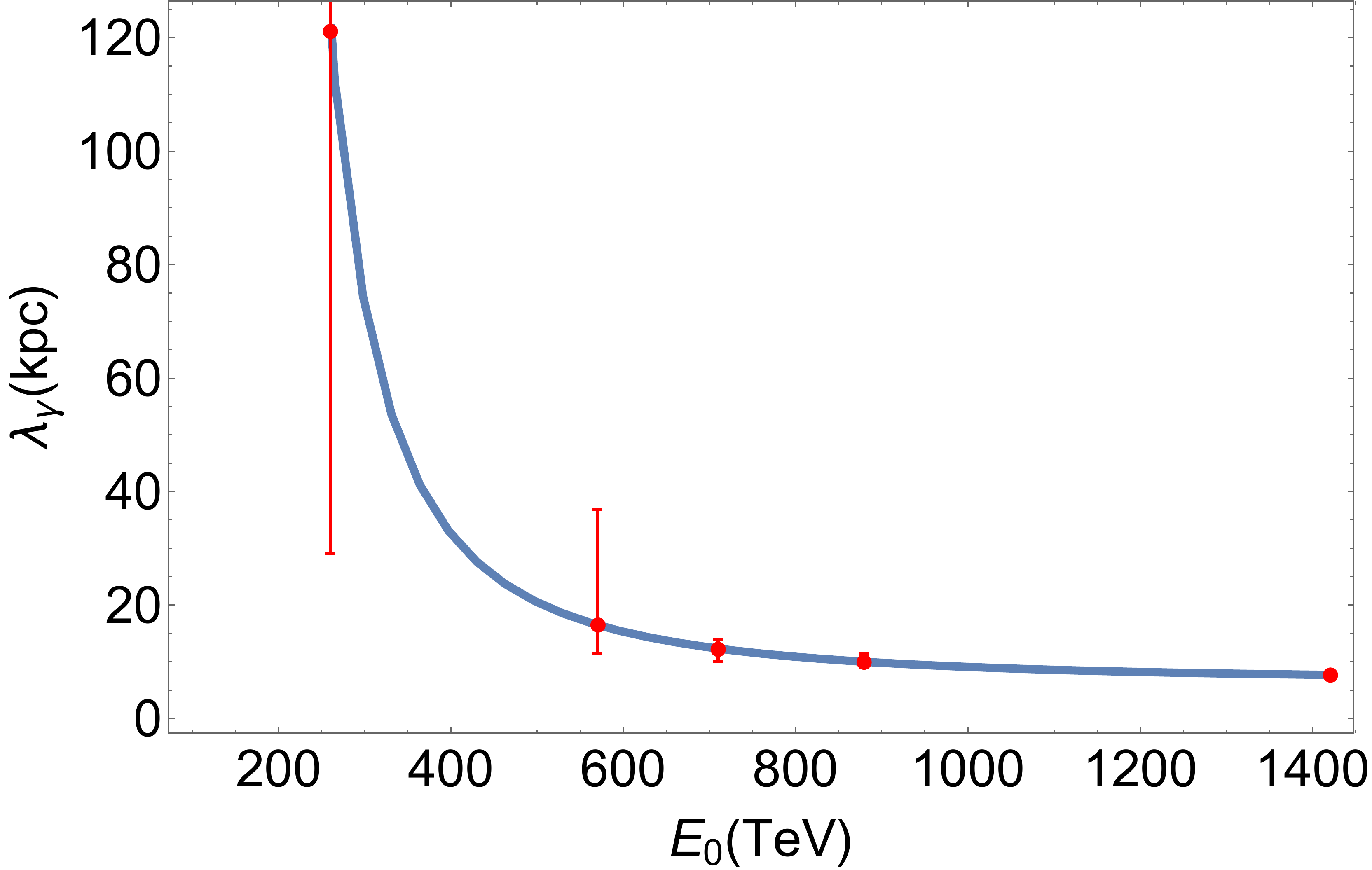}\ \hspace{0.05cm}
  \caption{\label{fig1} The relationship between the mean free path and the energy of observed photons, where the LHAASO detected gamma ray photons are marked by red dots.}}
\end{figure}

\begin{figure} [h]
  \center{
  \includegraphics[scale=0.25]{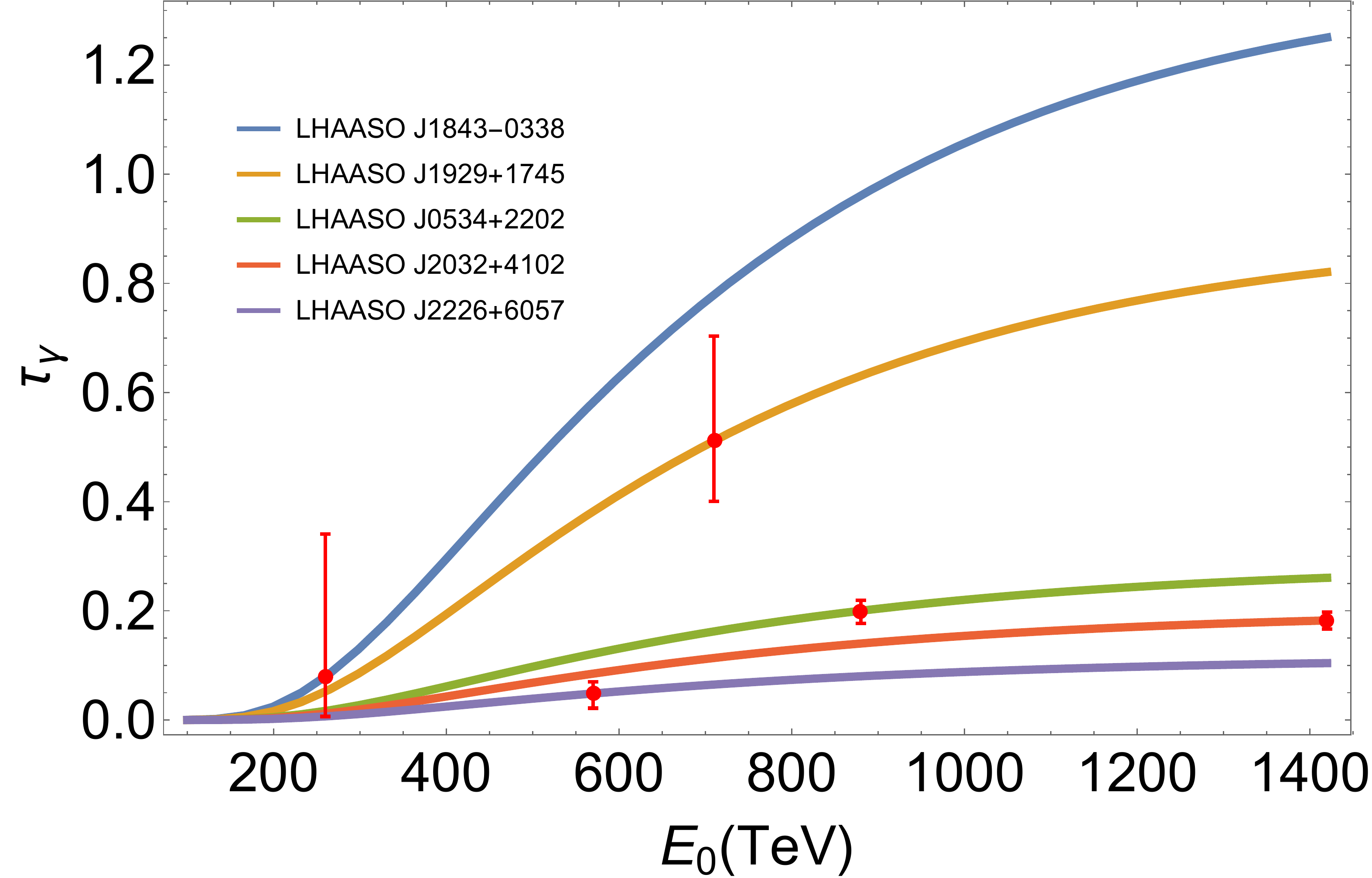}\ \hspace{0.05cm}
 \includegraphics[scale=0.25]{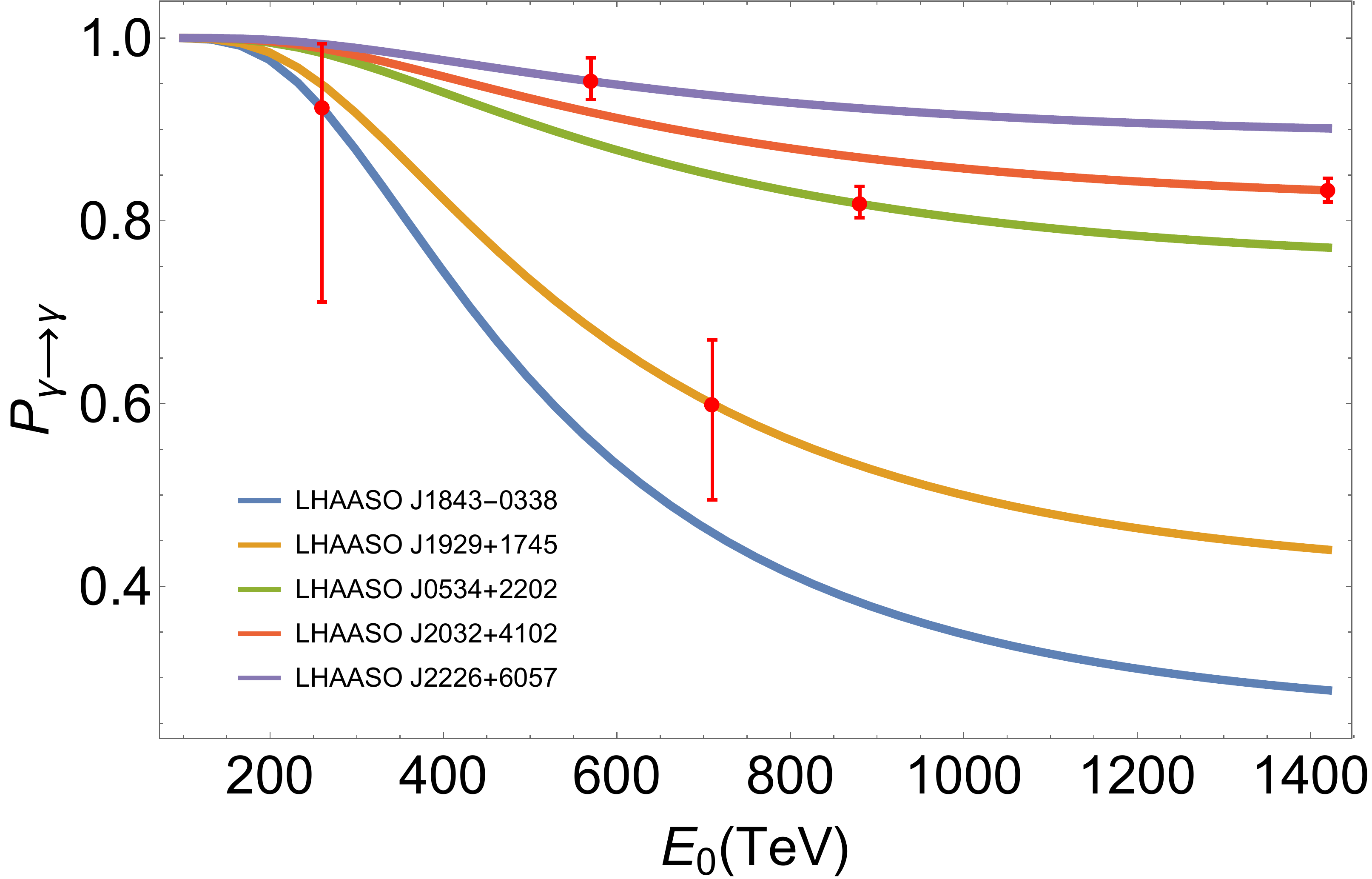}\ \hspace{0.05cm}
  \caption{\label{fig2} The optical depth (left) and the survival probability (right) as the function of energy, where the LHAASO detected gamma ray photons are marked by red dots.}}
\end{figure}

In this section we apply the above theoretical analysis to LHAASO
data. The optical depth, mean free path and the survival
probability are computed for the eleven LHAASO sources and the
main results are summarized in Tab.\ref{Tabel1}.  While to
demonstrate the relations of these quantities explicitly,  we
choose five LHAASO sources to depict in each figure, four of which
(LHAASO J2032+4102, J0534+2202, J1843-0338 and J2226+6057) have
definite distance \footnote{In the sense of ignoring the
statistical uncertainties, these sources have definite
distances.}, while for the remaining one (LHAASO J1929+1745), we
choose the possible location with the maximal distance $6.3
\,\text{kpc}$.  Firstly, we demonstrate the relationship between
the mean free path and the observed energy for different LHAASO
sources, as illustrated in Fig.\ref{fig1}. Without surprise, it
indicates that the mean free path keeps decreasing as the energy
increases, since the gamma ray photons interacts with CMB photons
more easily. In addition, we plot the optical depth and survival
probability as the function of energy for different LHAASO
sources, as illustrated in Fig.\ref{fig2}. In general, the optical
depth increases with the energy of gamma ray photons. It is also
evident that the survival probability becomes lower with the
increase of the energy of gamma-ray photons. Nevertheless, as a
whole, even taking the statistical uncertainties into
account we find the survival probability stays at a very high
level, which  can be seen directly from Tab.~\ref{Tabel1}.
This result
 is qualitatively the same as depicted in Fig.6 of Ref.\cite{Cao}.

\begin{figure} [h]
  \center{
  \includegraphics[scale=0.22]{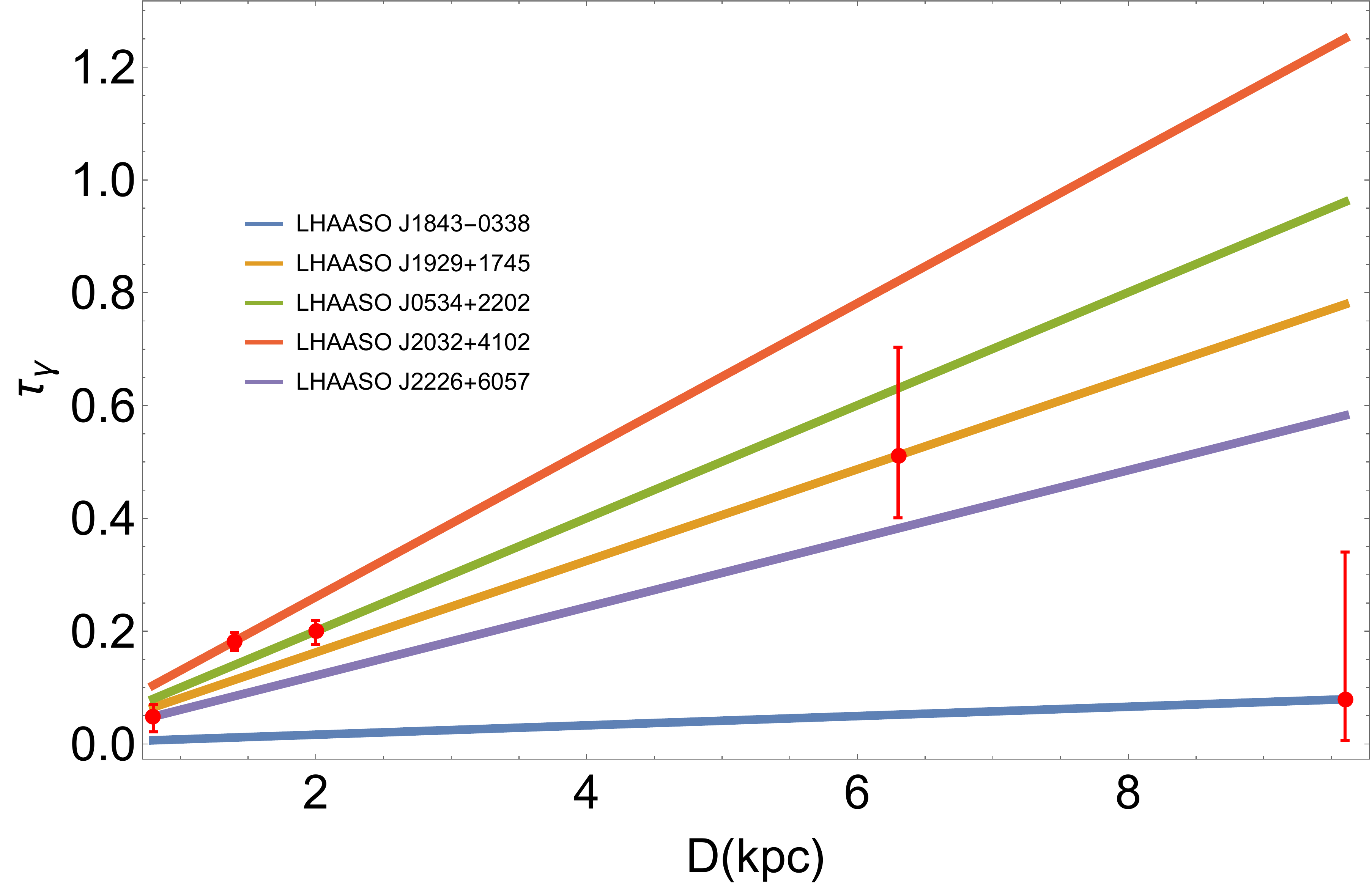}\ \hspace{0.05cm}
 \includegraphics[scale=0.23]{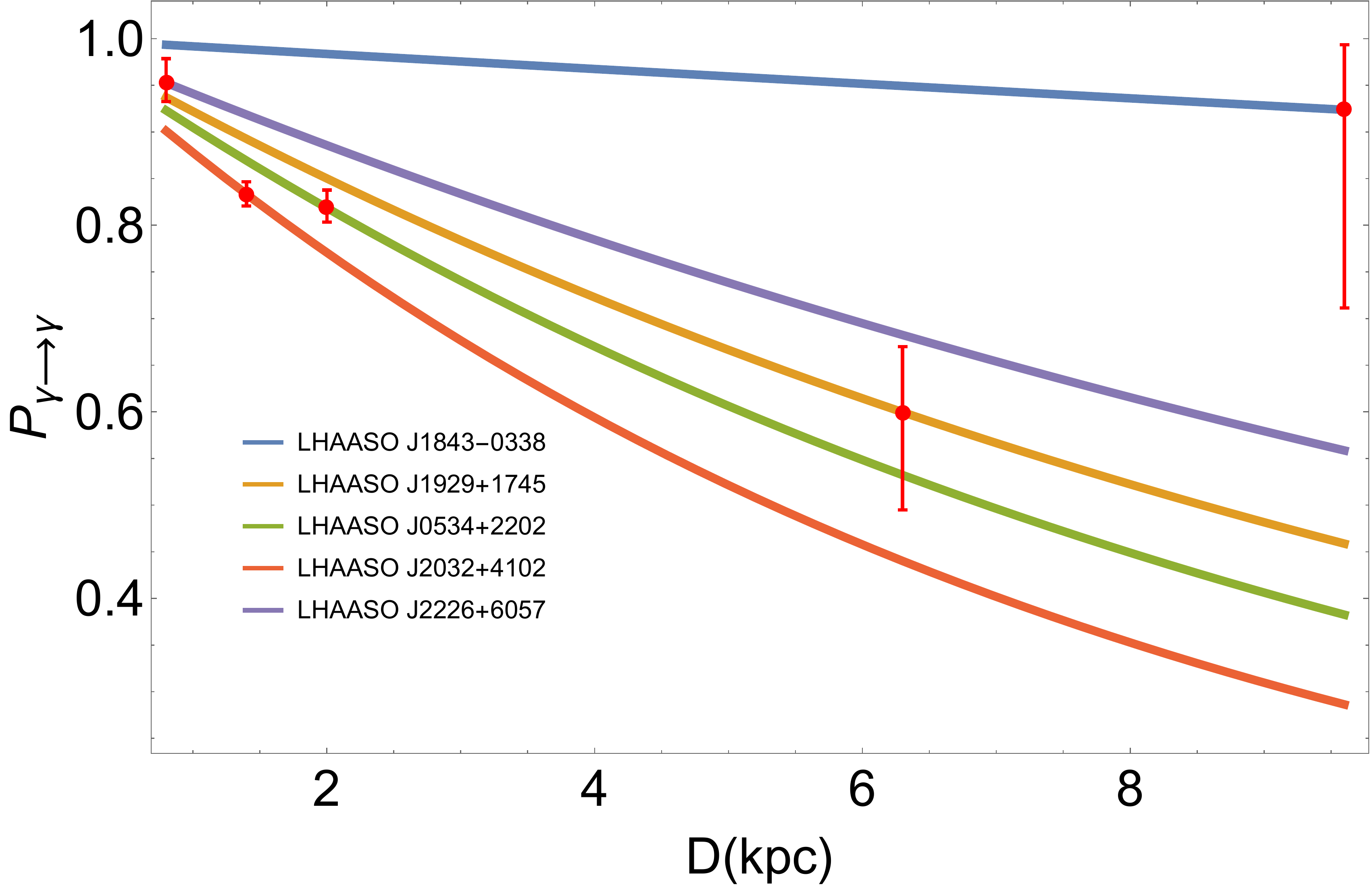}\ \hspace{0.05cm}
  \caption{\label{fig4} The optical depth (left) and the survival probability (right) as the function of source distance, where the LHAASO detected gamma ray photons are marked by red dots.}}
\end{figure}

In parallel, we  plot the optical depth and the survival
probability as the function of the source distance, as shown in
Fig.\ref{fig4}. We notice that the optical depth increases with the increase of the source distance, while the survival probability decreases. This trend is reasonable because the UHE gamma rays from the source farther from the Earth are more likely to interact with the background.  In addition, we remark that  for galactic sources,  the mean free path is not influenced by the source distance to the leading order of $D$, which is evident if one substitutes (\ref{eq_od}) into (\ref{eq_mfp}). It indicates that the mean free path is only sensitive to the energy for galactic sources in this context.

Finally, we turn to the main results summarized in Tab.\ref{Tabel1}, which lists the optical depth, mean free path as well as the
survival probability for the eleven LHAASO sources. First of all,
from the fourth column in Tab.\ref{Tabel1} we notice that the
optical depth for all the LHAASO sources is much less than one,
namely $\tau_{\gamma} \ll 1$, which manifestly indicates that a large
amount of gamma-ray photons emitted from the source can reach the
Earth without being absorbed. Furthermore, from the last column in
Tab.\ref{Tabel1} we notice that most of the survival probability
for gamma ray photons is above 0.8. For instance, Ref.\cite{Cao}
observes that the gamma rays in LHAASO J2032+4102 have energy of
1.42 PeV, and the survival probability of such high-energy photons
reaching the Earth is $P_{\gamma \rightarrow \gamma}\left(E_{0},
z_{s}\right)\simeq83.3\%$. The lowest survival probability for
observed gamma ray photons comes from LHAASO J1929+1745, which is
about 0.60. This further shows that although the energy of gamma
ray photons is 0.71 PeV and beyond the threshold value of pair
production, a large number of gamma-ray photons can still reach
the Earth without the violation of Lorentz symmetry.

Therefore, we intend to conclude that it is still far to argue
that the Lorentz symmetry would be violated due to the present
observations from LHAASO.

\section{Discussion}\label{sec3}
In this note we have computed the optical depth, the mean free
path as well as the survival probability for photons from all the
gamma-ray sources detected by LHAASO in Ref. \cite{Cao}. The cross-section of the
pair production due to the interaction with CMB photons is
obtained within the standard special relativity and we find the
survival probability is fairly high for galactic gamma ray
photons, even though the energy of those photons may exceed the
threshold value of pair production. Thus there is no tension to
consider the violation of Lorentz symmetry. This should be true
for general galactic pevatrons, because in comparison with the
cosmic scale, the distance between the galactic sources and the
Earth is still too close to provide enough chances to collide with
CMB photons during the propagation.

To be more specific, the current data from \cite{Cao} are
not sufficient to result in a subluminal correction constrained by
the pair-production reaction $\gamma \gamma \rightarrow e^{+}
e^{-}$. While the superluminal corrections constrained by the
photon decay reaction does not conflict with our results,
since the distances of possible origins shown in \cite{Cao} are
far enough for photons to decay \cite{Martinez-Huerta:2016odc}.

Nevertheless, we may wonder what kind of observations on UHE
cosmic rays would imply that one is urged to consider the Lorentz
symmetry violation. For gamma-ray photons, such kind of condition
would be reached once the optical depth is much close to one or
larger than one. From Fig.\ref{fig4}, as suggested also in \cite{Li:2021cdz,Li:2021tcw}, one would expect that if
some of PeV photons with much higher source distance would be detected
by LHAASO in future, then the tension of violating Lorentz
symmetry would become strong.

Of course, it is completely possible to consider the corrections
of the survival probability due to the Lorentz symmetry violation. For
instance, one may modify the dispersion relations for photons and
obtain the threshold energy with corrections, and finally plot the
optical depth with Lorentz symmetry violation, as performed
in Ref. \cite{GuedesLang:2017sfl}.

\vspace{3mm}
\centerline{\rule{90mm}{0.9pt}}


\begin{thebibliography}{10}

\vspace{3mm}

\bibitem{Cao}
Cao, Z., Aharonian, F.A., An, Q. et al,
 Nature 594, 33-36 (2021).


\bibitem{Amenomori:2019rjd}
M.~Amenomori, Y.~W.~Bao, X.~J.~Bi, D.~Chen, T.~L.~Chen, W.~Y.~Chen, X.~Chen, Y.~Chen, Cirennima and S.~W.~Cui, \textit{et al.}
Phys. Rev. Lett. \textbf{123}, no.5, 051101 (2019)
[arXiv:1906.05521 [astro-ph.HE]].

\bibitem{TibetASgamma:2021tpz}
M.~Amenomori \textit{et al.} [Tibet ASgamma],
Phys. Rev. Lett. \textbf{126}, no.14, 141101 (2021)
[arXiv:2104.05181 [astro-ph.HE]].

\bibitem{HAWC:2019xhp}
A.~U.~Abeysekara \textit{et al.} [HAWC],
Astrophys. J. \textbf{881}, 134 (2019)
[arXiv:1905.12518 [astro-ph.HE]].




\bibitem{HAWC:2019tcx}
A.~U.~Abeysekara \textit{et al.} [HAWC],
Phys. Rev. Lett. \textbf{124}, no.2, 021102 (2020)
[arXiv:1909.08609 [astro-ph.HE]].


\bibitem{Carpet-3:2021omd}
D.~D.~Dzhappuev \textit{et al.} [Carpet\textendash{}3],
[arXiv:2105.07242 [astro-ph.HE]].

\bibitem{Hillas1984}
Hillas, Anthony M.
Annual review of astronomy and astrophysics 22.1 (1984): 425-444.




\bibitem{Aloisio2018}
Aloisio, R., Coccia, E.  Vissani, F. (eds) Multiple Messengers and Challenges in Astroparticle Physics (Springer, 2018).







\bibitem{Maccione:2010sv}
L.~Maccione, S.~Liberati and G.~Sigl,
Phys. Rev. Lett. \textbf{105}, 021101 (2010)
[arXiv:1003.5468 [astro-ph.HE]].







\bibitem{Liberati:2013xla}
S.~Liberati,
Class. Quant. Grav. \textbf{30}, 133001 (2013)
[arXiv:1304.5795 [gr-qc]].



\bibitem{Vasileiou:2013vra}
V.~Vasileiou, A.~Jacholkowska, F.~Piron, J.~Bolmont, C.~Couturier, J.~Granot, F.~W.~Stecker, J.~Cohen-Tanugi and F.~Longo,
Phys. Rev. D \textbf{87}, no.12, 122001 (2013)
[arXiv:1305.3463 [astro-ph.HE]].



\bibitem{Mavromatos:2010pk}
N.~E.~Mavromatos,
Int. J. Mod. Phys. A \textbf{25}, 5409-5485 (2010)
[arXiv:1010.5354 [hep-th]].



\bibitem{Shao:2010wk}
L.~Shao and B.~Q.~Ma,
Mod. Phys. Lett. A \textbf{25}, 3251-3266 (2010)
[arXiv:1007.2269 [hep-ph]].



\bibitem{Amelino-Camelia:1996bln}
G.~Amelino-Camelia, J.~R.~Ellis, N.~E.~Mavromatos and D.~V.~Nanopoulos,
Int. J. Mod. Phys. A \textbf{12}, 607-624 (1997)
[arXiv:hep-th/9605211 [hep-th]].



\bibitem{Amelino-Camelia:1998bln}
Amelino-Camelia, G., Ellis, J., Mavromatos, N. et al.
Nature 393, 763-765 (1998).

\bibitem{Mattingly:2005re}
D.~Mattingly,
Living Rev. Rel. \textbf{8}, 5 (2005)
[arXiv:gr-qc/0502097 [gr-qc]].



\bibitem{Jacobson:2005bg}
T.~Jacobson, S.~Liberati and D.~Mattingly,
Annals Phys. \textbf{321}, 150-196 (2006)
[arXiv:astro-ph/0505267 [astro-ph]].


\bibitem{Amelino-Camelia:2008aez}
G.~Amelino-Camelia,
Living Rev. Rel. \textbf{16}, 5 (2013)
[arXiv:0806.0339 [gr-qc]].


\bibitem{Liberati:2009pf}
S.~Liberati and L.~Maccione,
Ann. Rev. Nucl. Part. Sci. \textbf{59}, 245-267 (2009)
[arXiv:0906.0681 [astro-ph.HE]].







\bibitem{Chen:2021hen}
L.~Chen, Z.~Xiong, C.~Li, S.~Chen and H.~He,
Chin. Phys. C \textbf{45}, 105105 (2021)
[arXiv:2105.07927 [astro-ph.HE]].

\bibitem{Li:2021cdz}
H.~Li and B.~Q.~Ma,
JHEAp \textbf{32}, 1-5 (2021)
[arXiv:2105.06647 [hep-ph]].




\bibitem{Li:2021tcw}
C.~Li and B.~Q.~Ma,
Phys. Rev. D \textbf{104}, no.6, 063012 (2021)
[arXiv:2105.07967 [astro-ph.HE]].



\bibitem{Satunin:2021vfx}
P.~Satunin,
Eur. Phys. J. C \textbf{81}, 750 (2021)
[arXiv:2106.06393 [hep-ph]].




\bibitem{LHAASO:2021opi}
Z.~Cao \textit{et al.} [LHAASO],
[arXiv:2106.12350 [astro-ph.HE]].

\bibitem{Martinez-Huerta:2017ulw}
H.~Mart\'\i{}nez-Huerta and A.~P\'erez-Lorenzana,
Phys. Rev. D \textbf{95}, no.6, 063001 (2017)
[arXiv:1610.00047 [astro-ph.HE]].

\bibitem{Astapov:2019xmt}
K.~Astapov, D.~Kirpichnikov and P.~Satunin,
JCAP \textbf{04}, 054 (2019)
[arXiv:1903.08464 [hep-ph]].

\bibitem{Satunin:2019gsl}
P.~Satunin,
Eur. Phys. J. C \textbf{79}, no.12, 1011 (2019)
[arXiv:1906.08221 [astro-ph.HE]].

\bibitem{HAWC:2019gui}
A.~Albert \textit{et al.} [HAWC],
Phys. Rev. Lett. \textbf{124}, no.13, 131101 (2020)
[arXiv:1911.08070 [astro-ph.HE]].

\bibitem{Wei:2021ite}
J.~J.~Wei and X.~F.~Wu,
[arXiv:2111.02029 [astro-ph.HE]].





\bibitem{Gould:1967zza}
R.~J.~Gould and G.~P.~Schreder,
Phys. Rev. \textbf{155}, 1408-1411 (1967)


\bibitem{Moskalenko:2005ng}
I.~V.~Moskalenko, T.~A.~Porter and A.~W.~Strong,
Astrophys. J. Lett. \textbf{640}, L155-L158 (2006)
[arXiv:astro-ph/0511149 [astro-ph]].



\bibitem{Angelis}
A. De Angelis, G. Galanti, and M. Roncadelli, Monthly
Notices of the Royal Astronomical Society 432, 3245
(2013).

\bibitem{Martinez-Huerta:2016odc}
H.~Mart\'\i{}nez-Huerta and A.~P\'erez-Lorenzana,
J. Phys. Conf. Ser. \textbf{761}, no.1, 012035 (2016)
[arXiv:1609.07185 [astro-ph.HE]].

\bibitem{GuedesLang:2017sfl}
R.~Guedes Lang, H.~Mart\'\i{}nez-Huerta and V.~de Souza,
Astrophys. J. \textbf{853}, no.1, 23 (2018)
[arXiv:1701.04865 [astro-ph.HE]].


\end{thebibliography}
\end{document}